# Two energy gaps and Fermi surface "arcs" in NbSe$_2$

S. V. Borisenko[1], A. A. Kordyuk[1], V. B. Zabolotnyy[1], D. S. Inosov[1], D. Evtushinsky[1], B. Büchner[1], A. N. Yaresko[2], A. Varykhalov[3], R. Follath[3], W. Eberhardt[3], L. Patthey[4], H. Berger[5]

[1]*Leibniz-Institute for Solid State Research, IFW-Dresden, D-01171, Dresden, Germany*

[2]*Max-Planck-Institute for the Physics of Complex Systems, Dresden, Germany*

[3]*BESSY, Berlin, Germany*

[4]*Swiss Light Source, Paul Scherrer Institut, CH-5234 Villigen, Switzerland*

[5]*Institute of Physics of Complex Matter, EPFL, 1015 Lausanne, Switzerland*

(27 February 2008)

**Using angle-resolved photoemission spectroscopy (ARPES), we report on the direct observation of the energy gap in 2H-NbSe$_2$ caused by the charge-density waves (CDW). The gap opens in the regions of the momentum space connected by the CDW vectors, which implies a nesting mechanism of CDW formation. In remarkable analogy with the pseudogap in cuprates, the detected energy gap also exists in the normal state (T>T$_0$) where it breaks the Fermi surface into "arcs", it is non-monotonic as a function of temperature with a local minimum at the CDW transition temperature (T$_0$) and it forestalls the superconducting gap by excluding the nested portions of the Fermi surface from participating in superconductivity.**

Transition metal dichalcogenides are known for their rich phase diagrams which often include different types of electronic ordering. 2H-NbSe$_2$ exhibits the incommensurate charge density wave order below =33.5 K and becomes superconducting at T$_c$=7.2 K Refs.(1-3). Until recently, the underlying mechanism of the CDW in two-dimensional chalcogenides was not clear. In analogy with one-dimensional case, various types of nesting have always been in the focus of earlier and most recent studies[4-10], although no consensus has been reached. Common to all ARPES studies on 2H-NbSe$_2$ is the lack of a CDW induced energy gap being observed. The rather weak influence of the transition at T$_0$ on the transport properties[3,11] and optical spectra[12] is in line with this non-observation, but is in sharp contrast with a 35 meV energy gap seen in tunnelling spectroscopy [13,14]. This uncertainty in the experimental investigations was accompanied by a similarly broad spectrum of theoretical ideas, ranging from the nesting of the FS[15] or saddle points[10] to the nesting for CDW being irrelevant[16].

In a very recent study it was shown that in a similar in many aspects 2H-TaSe$_2$, both, the nesting and corresponding CDW gap do exist and it is the presence of the pseudogap in the normal state (above T$_0$) which hindered their direct observation in earlier experiments[17]. Since the pseudogap in 2H-TaSe$_2$ bears all spectroscopic signatures of the famous pseudogap in superconducting cuprates[18], it is important to investigate the related compound which not only exhibits CDW order but also becomes a superconductor at an accesible temperature. Furthermore, the study of energy gaps in 2H-NbSe$_2$ would shed some light on the controversy as for the co-existence of the CDW and superconductivity. According to the authors of Ref.9, the CDW in 2H-NbSe$_2$ does not result in energy gap and actually boosts the superconductivity, contrary to the conventional expectations.

In this Letter we demonstrate that the CDW gap not only can be directly observed in 2H-NbSe2 in particular regions of momentum space, but also it persists to the normal state, is nonmonotonic and competes with the superconductivity for the Fermi surface, as in the cuprates[18,19].

ARPES data were collected using the synchrotron radiation ("1$^3$-ARPES" facility at BESSY) within the range of photon energies (20 - 60 eV) from cleaved (001) surface of high quality single crystals. The overall energy and momentum resolutions were set to ~ 2 meV (8 meV) and to ~ 0.013 Å$^{-1}$ respectively for the low temperature measurements (Fermi surface mapping). In Fig.1 we show the result of a high-precision Fermi surface (FS) mapping in CDW state. The FS of 2H-NbSe$_2$ consists of double-walled "barrels" around the center (Γ-point) and corners (K-points) of the Brillouin zone (BZ) as well as the small three-dimensional "pancake" FS around the Γ-point which is in general agreement with recent experimental[6, 8, 9] and computational studies[6, 16, 20]. Having covered a large area in **k**-space with high momentum and energy resolutions (all data points shown in Fig.1 are original and no symmetrization has been applied) we are able to determine which FS regions are separated by the CDW vectors and which are not. A surprising new result is that none of the earlier proposed FS nesting scenarios would appear to be valid (see caption to Fig.1). However, there are regions of the FS that are interconnected exactly by the primary CDW vectors. Firstly, these are rounded corners of the inner K-barrels (yellow points in Fig.1). Closer inspection of the map (Fig.1) indeed implies that these parts of the FS appear to be missing (gapped): they are blurred and their intensity is lower. Additionally, these new "hot spots" seem to be connected by the primary vectors also with other $k_F$-points (see green and blue ones) of the outer Γ- and K-barrels. These observations are further corroborated by the autocorrelation map[17] (not shown) and real part of the calculated Lindhard function (Supplementary Information) that, though without sharp peaks, exhibit a large weight at the CDW vectors indicating the presence of relatively good nesting conditions of the FS in 2H-NbSe$_2$.

In order to check whether the elusive CDW energy gap is present in these regions, in Fig. 2a we show the distribution of photoemission intensity along the K-M-K boundary of the BZ in the CDW state. The energy gap can be determined as the difference between the binding energies of the leading-edge midpoints (LEM) of two, gapped and non-gapped, energy-distribution curves (EDC). As a reference EDC one usually takes the EDC corresponding to a Fermi momentum ($k_F$) where the energy gap is zero (see Supplementary Information). Here, in analogy with 2H-TaSe$_2$ (Ref. 17), we consider the $k_F$, closest to the M-point, i.e. the point on the outer K-barrel, as such a "node" (see also Fig.1). Thus, the cut along K-M-K line shown in Fig. 2a is ideal for such measurements: it contains two "nodes" and two "hot spots". In Fig. 2b we plot the binding energies of LEM for all momentum values shown in Fig. 2a. The presence of an energy gap in the CDW state is obvious, appearing as the difference between LEM binding energies of $k_{M+}$ and $k_{K+}$ EDCs of about 2.4 meV (Fig. 2c). Interestingly, the energy gap is also observed in the normal state (T = 119 K) and as follows from the lower panel of Fig. 2c, it is approximately two times larger than at low temperatures. Since the spectroscopic signatures of both gaps are the same and similar to the pseudogaps in underdoped cuprates and 2H-TaSe$_2$, we can refer to both as "pseudogaps" keeping in mind that one of them is the normal state gap and the other is the incommensurate CDW gap. To clarify the behaviour in between we measured, in detail, the temperature dependence of the gap values. Fig. 2d shows a non-monotonic behaviour of the energy gap magnitude as a function of temperature. The minimum of the gap corresponds to the CDW transition temperature $T_0$. The presence of the gap above and below the critical temperature masks the transition into the CDW state. It is for this reason that a straightforward comparison of the EDCs just above and below $T_0$ reveals no evidence for a gap in this system. The apparent trend to increase below $T_0$ (Fig. 2d) and the magnitude of the LEM gap at 19.5 K (~2.4 meV) are in reasonable agreement with the classical BCS value at zero temperatures (5.1 meV). On the other hand, the trend to increase above $T_0$ is not expected in the light of general knowledge about the pseudogap in the cuprates. However, if the pseudogap is measured directly in the "hot spot" in underdoped BSCCO all results including the case of 2H-TaSe$_2$ are in qualitative agreement[17,18].

The momentum anisotropy of the gap can be conveniently and precisely checked if the LEM is determined for a number of near-$k_F$ points and all values are collected in a "map of gaps"[22]. Such a map is presented in Fig. 3. If we neglect small variations of LEM along most of the FS (see note in Supplementary Information), the regions where the gap is observable are indeed well localized in **k**-space. The "hot spots" are visible as islands of higher gap values breaking apart the inner K-barrel, which can be considered now as three "arcs" just as is found in underdoped cuprates where the hole-like barrel around ($\pi$, $\pi$) is divided by the pseudogap in four "arcs". This detailed search for a gap in the BZ contrasts our results with the known tunnelling data[13,14] even more sharply. It is unlikely that ARPES simply overlooks a large gap since it is able to resolve an order of magnitude smaller gap. We offer a simple explanation for this drastic discrepancy. The peak in the tunnelling spectrum on the occupied part is simply a saddle point singularity at 35 meV, located on the ΓK-line between the outer barrels (see also Ref. 9). The other peak in the unoccupied part is most likely the top of the band responsible for the 3D pancake-like FS centred at the Γ-point.

In Fig. 4 we present the superconductivity-related data. The anisotropy of the superconducting gap in 2H-NbSe$_2$ has been studied before[8,9,23]. For the most part our data are in general agreement with these results, however, the superior resolution has enabled us to reveal two further important correlations. The larger superconducting gaps correspond to narrower coherence peaks and it is largest where the CDW gap is smallest or zero, and vice versa. Panel **a** of Fig. 4 shows a classical BCS-like opening of the superconducting gap along the cut through point #1 (Fig. 4b) which, according to Fig. 3, corresponds to the section of the FS not influenced by the CDW gap. The superconducting gap is near its largest here (Fig. 4c). In contrast, the superconducting gap in "hot spots" (point #3) is zero (Fig. 4c). Thus, contrary to Ref. 9, these observations highlight the competing nature of CDW and superconductivity.

Our first observation of the CDW gap on the parts of the FS separated by the primary CDW vectors together with its temperature and momentum dependences, not only firmly support the nesting scenario of CDW in layered dichalcogenides but also strongly point to the pseudogap in cuprates being due to a density wave, generally incommensurate, and not arising from fluctuations of superconductivity. This follows from a remarkably complete analogy between the pseudogaps in two classes of compounds. The present data along with recent results on 2H-TaSe$_2$ (Ref. 17) emphasize how elusive and strange the signatures of the incommensurate density waves may appear in the ARPES spectra. In spite of the phase transition, the FS stays virtually the same. This is because the energy gap is not complete (pseudo) and is present above and below $T_0$. In the CDW state the pseudogap is strongly anisotropic in momentum space and if data analysis is not sensitive to its very small values (e.g. symmetrization), the FS may appear to be "broken in arcs". In reality, even though the spectral weight distribution in the incommensurate state is very specific, it represents an intermediate structure caused by the smooth transition from the original unreconstructed FS to the folded one, and all features have a transparent physical meaning. As in the case of 2H-TaSe$_2$ (Ref. 17) and, possibly, underdoped cuprates, the spectral weight of the folded FS replica is extremely small and therefore hardly visible in 2H-NbSe$_2$. In the superconducting state, as in underdoped cuprates[19], the superconducting gap acts only on those parts of the FS that are not (or not significantly) modified by the pseudogap at higher temperatures.

The long list of analogies between the pseudogap regime of underdoped cuprates and layered dichalcogenides (pressure and doping phase diagrams, resistivity[11], Nernst effect[24], optics[25], Hall effect[26], tunnelling[27, 28], "kinks" in dispersion[8], non-monotonic pseudogap and its anisotropy[17, 18], etc.) can now be extended to include FS "arcs" and a "two-gap scenario". It is known that the superlattice peaks in neutron scattering experiments are not universally present in all underdoped cuprates, but the pseudogap exists already in the normal state of dichalcogenides where these peaks are absent as well. Nevertheless, we do not claim that the pseudogap in cuprates is the consequence of a simple, classical charge density wave state. It is rather a general phenomenon, which may occur due to any type of incommensurate density ordering, including exotic ones[29]. We would also like to point out that the symmetry of the superconducting state in 2H-NbSe$_2$ is apparently different from the d-wave (or f-wave) that speaks in favour of a different underlying mechanism. At the very least we were not able to find distinct gapless regions on the Fermi surface similar to the nodal points in HTSCs. Regions with reduced superconducting gaps do exist and this is in agreement with the recent penetration depth measurements[30]. Density waves and superconductivity meet very often and thus, probably, not by accident. In their competition for the Fermi surface density waves seem to always be the winner – $T_0$ is always higher than $T_c$ (Ref. 31).

The project was supported, in part, by the DFG under Grant No. KN393/4 and BO 1912/2-1. We thank R. Hübel for technical support and M. Rümmeli for careful reading of the manuscript. This work has been supported by the Swiss National Foundation for Scientific Research and by the NCCR MaNEP.

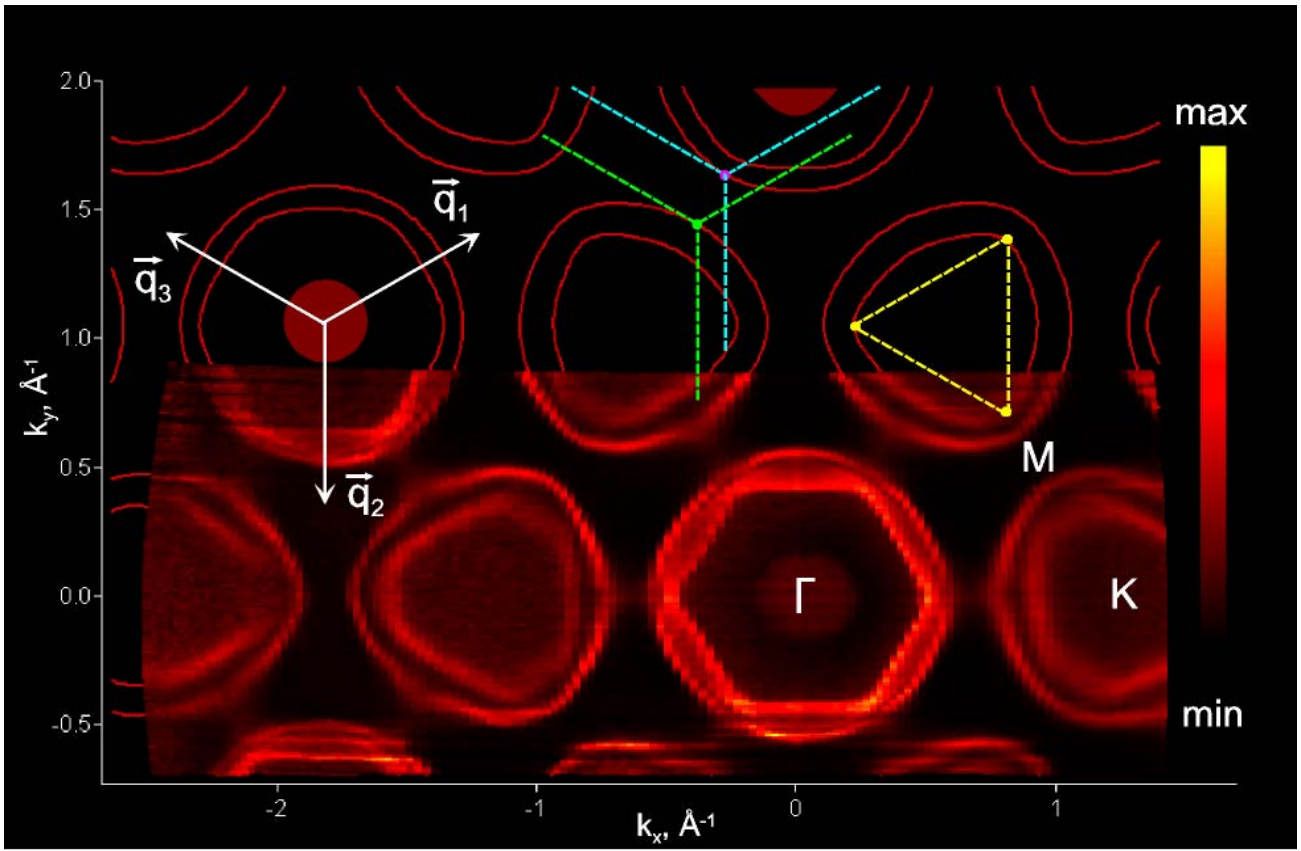

**Figure 1.** Fermi surface map of NbSe$_2$ at 20K (hv = 50 eV). Momentum distribution of the photoemission intensity at the Fermi level (14 meV window). The map is normalized to the integrated intensity (356 meV window) to enhance the FS-related features. Contours are fits to the experimental data. The white arrows and colour dashed lines are the primary CDW vectors. The Γ-centred barrels are far too large to be self-nested with the primary CDW vectors[4] and are smaller than would be expected for self-nesting with the secondary CDW vectors (e.g. **q$_1$**-**q$_2$**)[9]. The shape of the inner K-barrel is still far from being suitable for perfect self-nesting[6, 8, 21]. The **k**$_F$-points on the K-barrel lying on the ΓK symmetry lines cannot be connected by the noticeably shorter CDW vectors (green dashed lines), contrary to that suggested in Ref.(9).

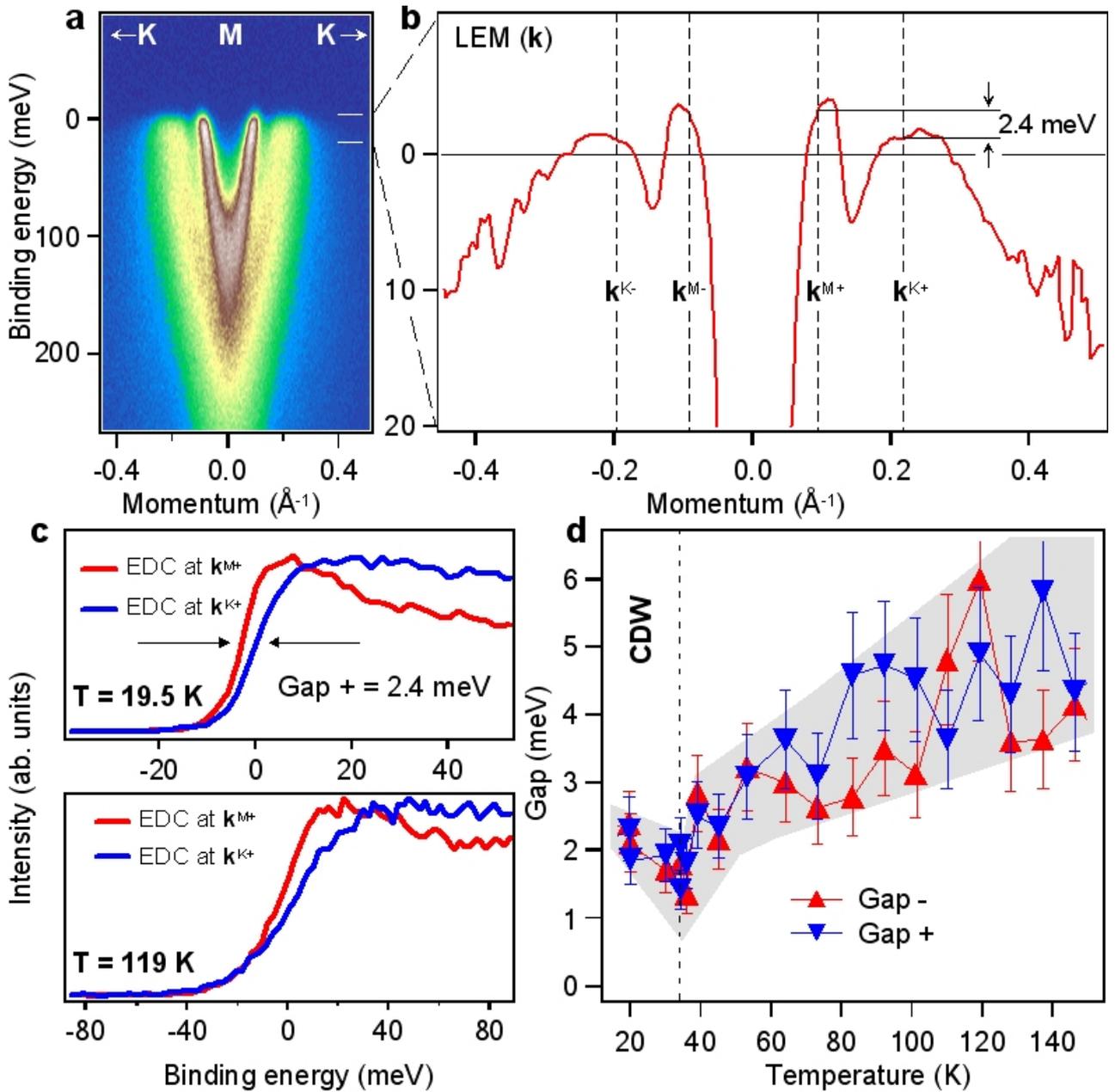

**Figure 2.** Energy gaps. a) Energy-momentum distribution of the intensity along the K-M-K high symmetry direction. (T= 19.5 K, hν = 51 eV) b) Leading edge midpoint (LEM) binding energy of all EDCs from **a**. The energy position of the $k_F$-EDC's leading edge was with an accuracy of better than 1 meV by taking a derivative of the fitting function. c) Direct evidence for energy gap in CDW (19.5 K) and normal state (119 K). d) Temperature dependence of the gap corresponding to positive and negative momentum values from **a**.

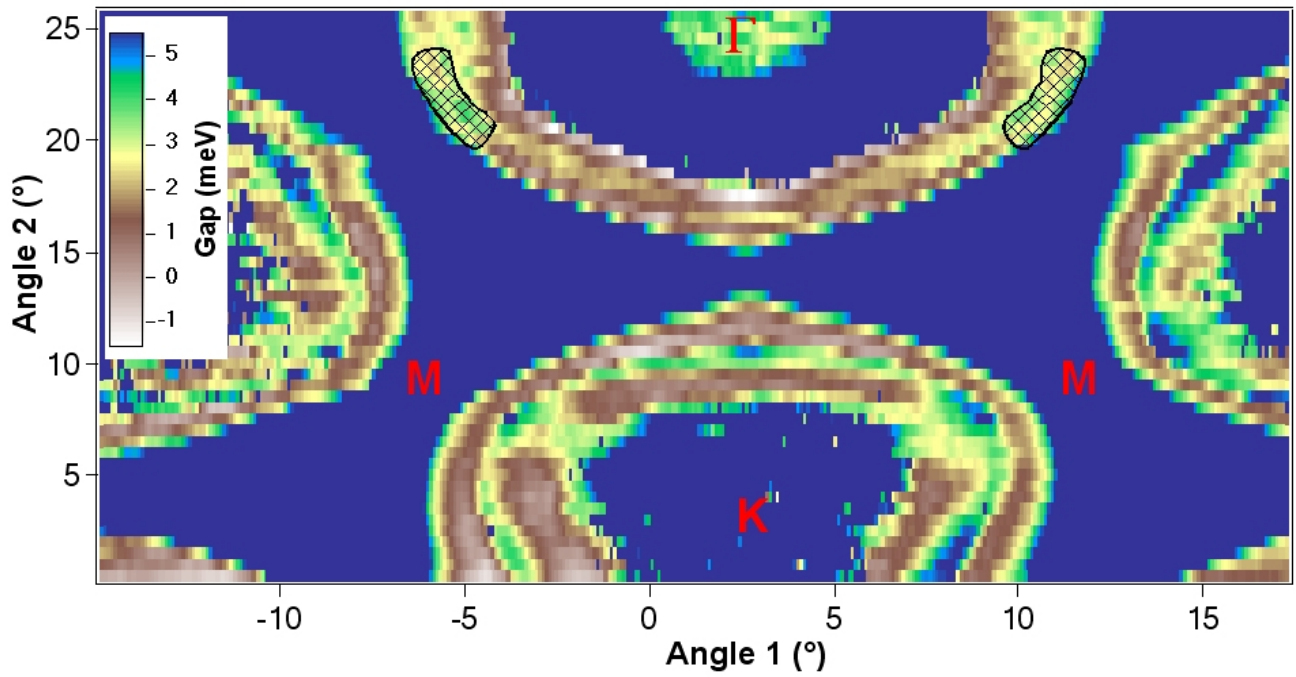

**Figure 3.** Fermi surface "arcs". Gap is defined as a difference between the LEM's binding energies of the $k_F$-EDCs and plotted as a function of **k** in corresponding angular space. Dashed areas designate the regions where the LEMs cannot be correctly defined. Fermi surface "arcs" are seen as disconnected segments of the inner K-barrel. Photon energy is 50 eV.

3333

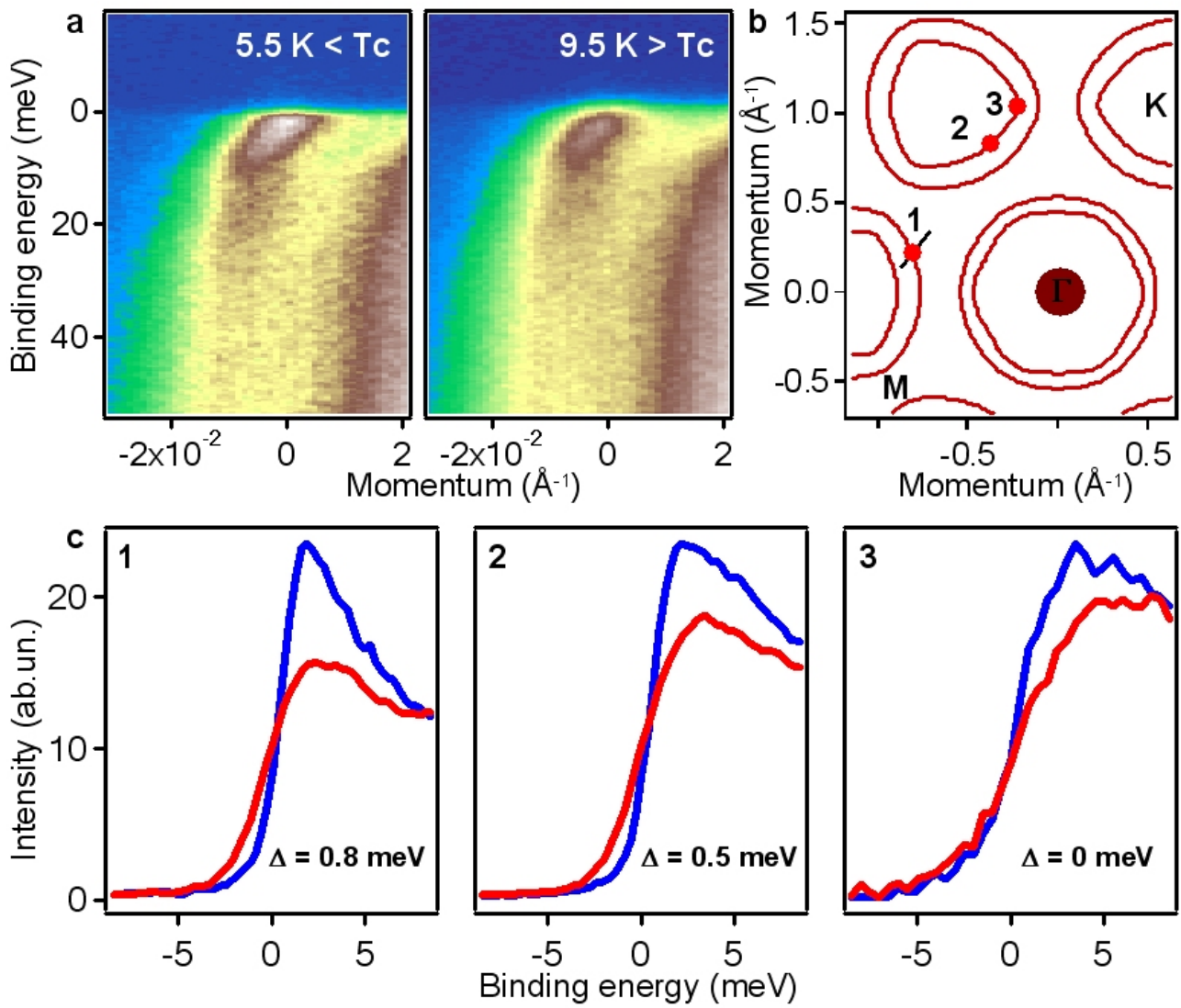

**Figure 4.** Superconductivity. a) Energy-momentum distribution of the intensity along the cut through the $k_F$-point #1 shown in b) below and above the critical temperature ($T_c$=7.2 K). Photon energy is 21 eV. c) $k_F$-EDCs from the points #1, #2 and #3 located as shown in b) below (blue) and above (red) the critical temperature $T_c$.

**Supplementary information**

There are many factors which may influence the position of the LEM [1s]. Our choice of reference EDC rules out the majority of them: Fermi velocity, temperature, energy and momentum resolutions. One may notice that the band supporting the outer K-barrel is weaker in terms of intensity. We have carefully studied the possible importance of the matrix element effects by varying the photon energy and thus relative intensities of these two features. The LEM as a function of momentum remains virtually intact (see Figure S1) in spite of the significant variation of the relative intensities. Finally, an additional broadening which is most likely caused by finite undulations of the inner K-centred FS (in reality the system is slightly 3D), cannot result in such a shift of the LEM according to both, simple simulations (not shown) and typical values of LEM corresponding to the inner Γ-centred barrel which reveals even stronger [9] undulations (Fig. 3).

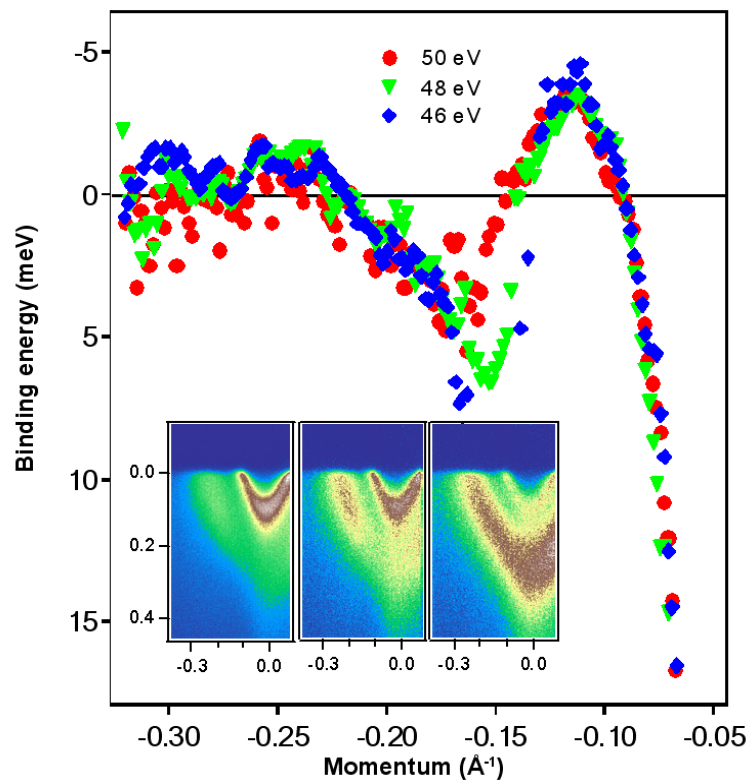

**Figure S1.** LEM measured using different excitation energies. Data are taken at Swiss Light Source.

Two remarks are necessary referring to Fig.3. First, what appears on the map as FS (see also Fig. 1) is a minimum LEM locus. It generally encloses a slightly smaller unoccupied **k**- space area than the true Fermi momenta usually do, nonetheless the momentum dependence of the gap is still quantitatively similar to that extracted from real $\mathbf{k}_F$ (see Fig. 2b). Secondly, in the rare case where two Fermi level crossings with a particular ratio of intensities are very close to each other in **k**-space, the typical behaviour of LEM(k) for one of the crossings can be distorted by the other, imitating the opening of the gap. For instance, the four crossings discussed earlier in Fig.2 a,b are quite well separated and the gap extraction is not sensitive to the mentioned effect, whereas along the ΓK direction the inner Γ-barrel comes very close to the outer one and results in an overestimated gap value (hatched areas in Fig. 3).

Fig. 3 also implies that there are **k**-points where the LEM protrudes even further above the Fermi level than the reference EDC. Strictly speaking this means that small gaps can also be present on other parts of the FS. This is however beyond the sensitivity of our experiment, because, as mentioned above, there are other factors which may be responsible for small variations of the LEM.

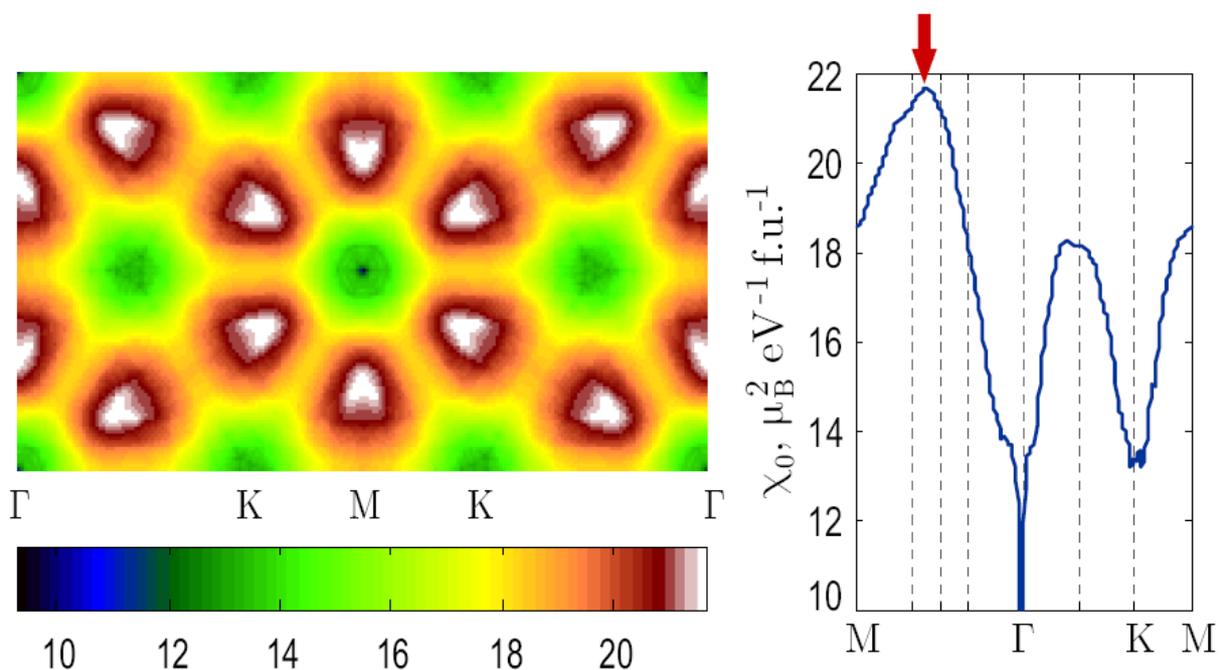

**Figure S2.** Real part of the Lindhard function at ω → 0 as a function of momentum and corresponding profile along high-symmetry directions, with the dominant nesting vector marked by the red arrow. The same vectors can be seen in left panel as white spots.

The nesting properties of NbSe2 are characterized by calculated from the experimental data static susceptibility (Figure S2). The peak at the incommensurate vector close to 2/3ΓM confirms the presence of the reasonable nesting in NbSe$_2$.